\documentclass[prd,twocolumn,superscriptaddress,showpacs,aps,amssymb]{revtex4}
\usepackage{graphicx} 
\usepackage{bm} 
\newcommand{\beq}{\begin{equation}}
\newcommand{\eeq}{\end{equation}}
\newcommand{\ba}{\begin{array}}
\newcommand{\ea}{\end{array}}
\newcommand{\lsim}   {\mathrel{\mathop{\kern 0pt \rlap
  {\raise.2ex\hbox{$<$}}}
  \lower.9ex\hbox{\kern-.190em $\sim$}}}
\newcommand{\gsim}   {\mathrel{\mathop{\kern 0pt \rlap
  {\raise.2ex\hbox{$>$}}}
\lower.9ex\hbox{\kern-.190em $\sim$}}}
\newcommand{\etal}{\emph{et al.}}

\begin{document}

\title{On the status of superheavy dark matter}

\author{R.~Aloisio}
\affiliation{INFN, Laboratori Nazionali del Gran Sasso, I-67010
  Assergi (AQ), Italy}
\author{V.~Berezinsky}
\affiliation{INFN, Laboratori Nazionali del Gran Sasso, I-67010
  Assergi (AQ), Italy}
\affiliation{Institute for Nuclear Research of the RAS, Moscow, Russia}
 \author{M.~Kachelrie\ss}
\affiliation{Institut for fysikk, NTNU Trondheim, N-7491,
  Trondheim, Norway}

\date{April 13, 2006}

\begin{abstract}
Superheavy particles are a natural candidate for the dark matter in
the universe and our galaxy, because they are produced generically
during inflation in cosmologically interesting amounts. 
The most attractive model for the origin of superheavy dark matter 
(SHDM) is gravitational production at the end of inflation. 
The observed cosmological density of dark matter determines the mass
of the SHDM particle as $m_X=$~(a few)$\:\times 10^{13}$~GeV, 
promoting it to a natural
candidate for the  source of the observed ultra-high energy cosmic rays 
(UHECR). After a review of the theoretical aspects of SHDM, we up-date
its predictions for UHECR observations: no GZK cutoff, flat energy
spectrum with $dN/dE\approx 1/E^{1.9}$, photon dominance and galactic
anisotropy.  We analyze the existing data and conclude that SDHM as 
explanation for the observed UHECRs is at present disfavored
but not yet excluded. We calculate the anisotropy relevant for 
future Auger observations that should be the conclusive test for this 
model. Finally, we emphasize that negative results of
searches for SHDM in UHECR do not disfavor SHDM as a dark matter
candidate. Therefore, UHECRs produced by SHDM decays and with the
signatures as described should be searched for in the future as
subdominant effect. 
\end{abstract}
\maketitle

\section{Introduction}  
\label{sec:introduction}
Superheavy Dark Matter (SHDM) is an interesting aspect of modern
particle physics and cosmology. Being first suggested as
explanation~\cite{BKV,KR} 
for the observation of cosmic rays with energy above the so-called
Greisen-Zatsepin-Kuzmin (GZK) cutoff, it has been
later developed into the study of a new form of dark matter. 

The concept of dark matter composed by superheavy particles with
$m_X \gsim 10^{13}$~GeV (further on we shall call them $X$ particles)  
at first glance seems to be exotic, mainly because of two
questions: How can SHDM particles have a lifetime exceeding the age of
the universe, and why their abundance should be dominant  
in the universe today? 
The answers to both questions have been known for a long time and are
based on quite general theoretical concepts.
The problem of the stability or the quasi-stability of a heavy
particle exists also for the lightest supersymmetric particle,
e.g. the neutralino. Discrete gauge symmetries protecting
neutralinos from fast decays work equally well for other particles. A
particular example of a particle with lifetime exceeding the age of
the universe was given in Ref.~\cite{crypton}. The large abundance of
superheavy relic particles can be provided by gravitational
production~\cite{ZS}, which works for superheavy particles very
similar to the production of density fluctuations during inflation.  

In contrast to usual thermal relics, like e.g. neutralinos, SHDM
particles are non-thermal relics and have been never in chemical
equilibrium with radiation. They must be produced very early, at the
end of inflation. Then it is enough to transfer a tiny fraction from
the energy of radiation to SHDM particles, less than $10^{-18}$, in
order to have $\Omega_X \sim 0.3$ now: The energy density of
non-relativistic $X$ particles diminishes with time as $1/a^3$, where
$a(t)$ is the scaling factor of the universe, while the energy density
of radiation diminishes as $1/a^4$.  When normalized at the
inflationary epoch, $a_i=1$, $a(t)$ has the enormous value
$\sim1\times 10^{22}$ now. Not surprisingly, this small energy fraction can
be transferred to $X$ particles by many different mechanisms, such
as thermal production at reheating \cite{BKV,ChKR-r}, the   
non-perturbative regime of a broad parametric resonance at preheating 
\cite{preheating}, and production by topological defects \cite{BKV,Kolb}. 

Thermal production of $X$ particles with $m_X \gsim 10^{13}$~GeV requires
a very high reheating temperature $T_{\rm rh}$. In supersymmetric
cosmology, $T_{\rm rh}$ is limited by gravitino overproduction
as $T_{\rm rh} \lsim (10^9 - 10^{10})$~GeV. However, the gravitino density 
can be diluted efficiently by entropy production during thermal 
inflation \cite{th-infl}. Thermal inflation solves the problem of 
overproduction of particle at reheating and allows higher temperatures 
with efficient SHDM production. In Ref.~\cite{Kolb}, it was shown
that the maximal temperature after inflation can be much higher than 
$T_{\rm rh}$.

$X$ particles are efficiently produced at preheating~\cite{preheating}.
This stage, predecessor of reheating, is caused by 
oscillations of the inflaton field after inflation near the minimum of 
its potential. Such an oscillating field can non-perturbatively 
(in the regime of a broad parametric resonance) produce intermediate 
bosons $\chi$, which  then decay into $X$ particles. The mass of the
$X$ particles can be one or two orders of magnitude larger than the inflaton
mass $m_{\phi}$, which should be about $10^{13}$~GeV. 

$X$ particles can be also produced by topological defects, such as
strings or textures. Particle production occurs at string
intersections or in collapsing texture knots. $X$ particles can also be
produced by hybrid topological defects, such as monopoles connected by
strings or walls bound by strings. The main contribution to 
the $X$ particle density is given by the earliest epochs, soon after
topological defect formation. Topological defects of the energy scale 
$\eta > m_X$ can be formed
in phase transitions at or slightly before the end of inflation. 
Efficient production of topological defects is predicted for the
preheating stage.  

However, the most remarkable creation mechanism for SHDM is its 
gravitational production~\cite{grav}. Particles are created
by time-variable gravitational fields during the expansion
of the universe. For this mechanism the interaction of $X$ particles
with other particles (e.g. with the inflaton) is not required, even
sterile particles are produced. The present abundance 
$\Omega_{\rm shdm}$ of the $X$ particles is mainly determined by its
mass $m_X$, while the dependence of $\Omega_{\rm shdm}$ on the reheating 
temperature $T_{\rm rh}$ is model-dependent. To
provide $\Omega_{\rm shdm}=0.27$, needed according to WMAP
observations \cite{WMAP}, the mass of the $X$ particle must be
(a few)$\,\times 10^{13}$~GeV.

The stability of $X$ particles can be ensured by discrete gauge symmetries. 
It must be weakly broken, if we want long-lived particles with
lifetime $\tau_X \gsim t_0$, where $t_0$ is the age of the universe.    
This superweak symmetry breaking can occur due to wormhole~\cite{BKV}
or instanton effects~\cite{KR}. Alternatively, discrete gauge symmetries
could be broken by higher-dimensional operators~\cite{Ha98}. An 
example of a SHDM particle in a semi-realistic particle physics model
are cryptons, i.e.\ bound-states from a strongly interacting hidden
sector of string/M theory~\cite{crypton,crypton1,Coriano}.  

What are the prospects to observe SHDM, if $X$ particles are absolutely
stable?  

This is a pessimistic case for SHDM, because 
unitarity limits severely the $XX$-annihilation cross section: Since the
velocity of $X$ particles is very small, $v \ll 1$, only the s-wave
contributes to $\langle\sigma_{\rm ann}v\rangle=a +bv^2+{\cal O}(v^4)$,
resulting in an unobservable small cross section, 
$\langle\sigma_{\rm ann}v\rangle \approx a\lsim 1/m_X^2$.  
An interesting and rather exceptional case was found in 
Ref.~\cite{Khlopov}, when 
$X$ particles are cosmologically produced  in the form of close pairs 
and form bound states due to gauge interactions between them. 
The lifetime of pairs corresponds to the spiral-in time of pairs 
with their subsequent annihilation, as in the case of 
monopole-antimonopole pairs. 

In the framework of gravitational production of SHDM, the mass
of the $X$ particle is fixed as (a few)$\:\times 10^{13}$~GeV, and we are
left in the SHDM model with only one free parameter, the lifetime of
the $X$ particles $\tau_X$ . If one requires that the ``AGASA excess''
is explained by the SHDM model (see below), then this parameter is
fixed by the UHECR flux observed by this experiment. 

At present, the most interesting manifestation of SHDM may be the
observations of UHECR beyond the GZK cutoff~\cite{BKV,KR,BS,BBV,DT,BM,TW}. 
There are three basic signatures of UHECRs from SHDM: 

\begin{itemize}
\item
SHDM particles as any other DM particles cluster gravitationally and
accumulate in the halo of our galaxy with an overdensity $2.1\times 10^5$.
Hence the UHECR flux from SHDM has no GZK cutoff in energy spectrum \cite{BKV}.
\item
Since in the decays or annihilations of $X$ particles  pions are more
abundant than nucleons, UHE neutrinos and photons are the dominant
component of the primary flux \cite{BKV}.  
\item
The non-central position of the Sun in the galactic halo results in an
anisotropic UHECR flux from SHDM~\cite{DT}. 
\end{itemize}

The quantitative predictions for the energy spectra and the photon/nucleon 
($\gamma/N$) ratio in the decays or annihilations of $X$ particles required 
the extension of existing QCD calculations for parton cascades from
the TeV scale up-to the scale $m_X$. The first calculations used the  
analytic limiting spectrum approximation \cite{BK-limsp} or extended 
the Monte Carlo simulation HERWIG \cite{BS}. More recent calculations
using a SUSY-QCD Monte Carlo \cite{BK-MC,ABK} and the DGLAP evolution
equations\cite{ST,BD,ABK} predict quite accurately the secondary
spectra  from decays/annihilation of SHDM particles and agree well
with each other. Their most important outcome for UHECR observations 
is the flat shape of the energy spectrum. 
At the relevant energies, it can be approximated as $dE/E^{1.9}$
while the photon/nucleon ratio is $\gamma/N =r_{\gamma/N}(x)\approx 2 - 3$
\cite{ABK}, being only weakly dependent on $x=2E/m_X$.  

An anisotropic UHECR flux from SHDM is guaranteed by the fact that the
distance from the Sun to the outer boundary of the DM halo is largest in
the direction of the galactic center (GC). Numerical simulations of
the DM distribution show an increase of the DM density towards the GC
as $\propto r^{-1}$~\cite{NFW}  
or $\propto r^{-1.5}$~\cite{Moore}. This further enhances the expected 
anisotropy. The relevant
calculations for this anisotropy were presented in Ref.~\cite{BBV}.
Comparisons of the calculated anisotropy \cite{BM,TW} with existing
data of the air-shower arrays on the northern hemisphere have revealed
no contradiction between data and model predictions. By contrast, detectors   
on the southern hemisphere able to observe the GC are much more 
sensitive to this anisotropy. The data of the old SUGAR detector
located in Australia are only marginally consistent with the
prediction of the SHDM model~\cite{KS}.    

At what energy UHECRs from SHDM become the dominant component? 

\begin{figure}[ht]
\includegraphics[width=8.0cm]{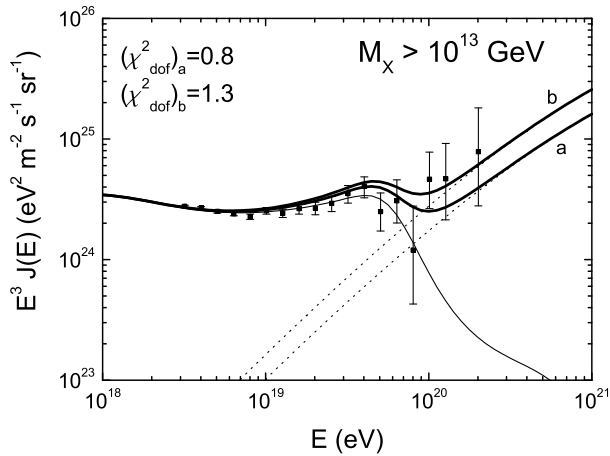}
\caption{The calculated spectrum of UHECRs from SHDM (dotted curves) in
comparison with the AGASA data (from Ref.~\cite{ABK}). The spectrum
from SHDM decays can explain only the highest energy events 
(``AGASA excess''). The dashed curve gives the
universal spectrum of extragalactic protons \cite{BGG}. The sum of
these two spectra is shown by full curves for two SHDM fluxes with different 
normalizations. The $\chi^2$ values are given for the comparison of these
curves with the experimental data at $E\geq 4\times 10^{19}$~eV. One can
see that the SHDM flux can dominate only at energy higher than 
$8\times 10^{19}$~eV.}
\label{shdm}
\end{figure}

The answer to this question became unambiguous after the precise
calculation of the spectrum of secondary particles produced in the
SUSY-QCD cascade, which can be approximated as 
$\propto E^{-1.9}$. This spectrum is very flat and
fitting it to the AGASA data~\cite{AGASA} shows that it can become the
dominant component of the UHECR flux only at energies above 
$8\times 10^{19}$~eV (see Fig.~\ref{shdm}). This is an
important and reliable conclusion about the status of UHECR from SHDM.

The photon dominance is another reliable prediction of the SHDM
model. Note that this test is relevant mainly for energies higher than
$8\times 10^{19}$~eV, where UHECRs from SHDM dominate the flux . Proton and
photon induced showers can be distinguished by the muon component
observed at ground level. An analysis of events with energies $E \geq
1\times 10^{20}$~eV has been performed for the AGASA data
\cite{AGASA-gamma,Risse} and very recently in Ref.~\cite{Rubtsov} for
the  combined AGASA \cite{AGASA} and Yakutsk data \cite{Ya}. We shall
discuss this analysis in Section 2. Here we only note that no
photon induced showers have been found among six AGASA  and four
Yakutsk events with energies higher than $1\times 10^{20}$~eV. One may
conclude therefore that the SHDM model is not confirmed by this analysis.  

We summarize our introduction emphasizing that at present only the
``AGASA excess'' at $E \gsim 1\times 10^{20}$~eV motivates the SHDM model 
as explanation for the existing UHECR data. The data of other detectors,
e.g. Yakutsk or HiRes, are compatible with the GZK cutoff. In this
case, the UHECR flux from SHDM can be only subdominant and is
reduced compared to the fit to the AGASA data shown in Fig.~1.
If the Auger experiment confirms the ``AGASA excess'', the status of
this model will be changed. Even in this case, the Auger detector has the
great potential to confirm or to reject the SHDM model for UHECRs by
testing its clearest signature,  the anisotropy towards the galactic
center. 

One must clearly emphasize the following. 

The observations of UHECR cannot exclude SHDM as explanation for the dark
matter in the universe and in our galaxy. Assuming that the SHDM are 
gravitationally produced, the mass of the $X$ particles is fixed and 
the only free parameter of the SHDM model is the life-time $\tau_X$. The 
``AGASA excess''  fixes this parameter as $\tau_X\approx 10^{20} $~years. 

>From the HiRes, Fly's Eye and Yakutsk data, that are compatible with
the GZK cutoff, only a lower bound on $\tau_X$ can be derived. Within
this lower bound, SHDM may still provide subdominant, but observable
effects in UHECR observations, and some of the showers observed at
the highest energies could be induced by secondaries from $X$ decays.   
Thus the search for photons  coming from directions close to the galactic
center remains an interesting task for the Auger detector, even if the
``AGASA excess'' will be not confirmed.

The discussion above determines the strategy of this paper. In 
Section~\ref{sec:gamma}, we 
obtain more accurately than in our previous work \cite{ABK} the ratio     
$\gamma/N$ for the SHDM model. Note, that the nucleon flux at 
$E \lsim 1\times 10^{20}$~eV is given mostly by extragalactic protons 
and thus our prediction is determined by the AGASA excess 
(see Fig.~\ref{shdm}) and valid only for the AGASA data. We compare 
our prediction with existing analyses of the $\gamma/p$ ratio using
the AGASA data. The Yakutsk or HiRes data require a larger value for
$\tau_X$ due to their agreement with the GZK cutoff and thus the $\gamma/p$
ratio derived for the AGASA data is in this case an {\em upper
limit\/}, not a prediction. In Section~\ref{anisotropy}, we calculate the 
anisotropy of UHECRs from SHDM relevant for Auger observations,
before we conclude.  

\section{Restrictions from photon-induced showers}
\label{sec:gamma}

In Ref.~\cite{ABK} we have calculated the ratio $r_{\gamma/N}(x)$ of
photons to nucleons in the QCD cascade initiated by the decay of a $X$
particle as function of $x=2E/m_X$.
Here we shall compute $\varepsilon_{\gamma}(E)= (\gamma/{\rm tot})_E$ as 
the ratio of photon-induced showers to the total number of showers 
at the measured energy $E$ using the following set of equations,
\beq
\varepsilon_{\gamma}(E)=
\frac{J_{\gamma}^{\rm shdm}(\lambda E)}
{J_p^{\rm shdm}(E)+J_p^{\rm extr}(E)+J_{\gamma}^{\rm shdm}(\lambda E)} ,
\label{obs}
\eeq
\beq
J_{\gamma}^{\rm shdm}(\lambda E)+J_p^{\rm shdm}(E)+J_p^{\rm extr}(E)=
J_{\rm AGASA}(E) ,
\label{AGASA}
\eeq
\beq
J_{\gamma}^{\rm shdm}(E)/J_p^{\rm shdm}(E)= r_{\gamma/N}(x) ,
\label{qcd}
\eeq
where the diffuse fluxes $J(E)$ with indices 'shdm' and 'extr' refer 
to SHDM and extragalactic fluxes, respectively. The SHDM fluxes are
taken as average over the galactic directions observed by AGASA. 
As extragalactic proton flux we use the universal spectrum from 
Ref.~\cite{BGG} as shown in 
Fig.~\ref{shdm}. The SHDM spectra $J_{\gamma}^{\rm shdm}(E)$
and $J_p^{\rm shdm}(E)$ are taken from Ref.~\cite{ABK} (not using the
power-law approximation). The photon flux  $J_{\gamma}$ is evaluated
at the energy $\lambda E$, where $E$ is the energy determined 
experimentally assuming that the primary is a proton. 
The coefficient $\lambda $ takes into account the differences in the
shower development between showers initiated by protons and by photons.
These differences are caused mainly by the Landau-Pomeranchuk-Migdal
effect.

Equation~(\ref{AGASA}) normalizes the total flux at 
$E\gsim 1\times 10^{20}$~eV to the observed ``AGASA excess'' 
(see Fig.~\ref{shdm}).  
\begin{figure}[ht]
\includegraphics[width=8.0cm]{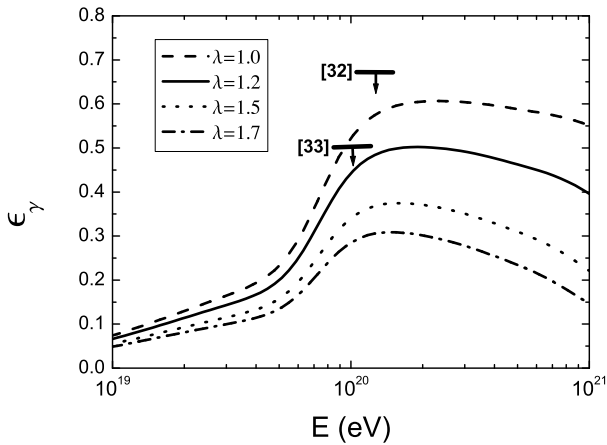}
\caption{Predicted ratio $\varepsilon_{\gamma}$ as function of the 
observed energy for $m_X=1\times 10^{13}$~GeV and for  
different values of the $\lambda$ parameter. The solid curve 
corresponds to $\lambda = 1.2$, valid according to 
Ref.~\cite{AGASA-gamma,Risse} for the AGASA site.
The two limits in the figure are those obtained in \cite{Risse} 
($\varepsilon=0.67$) and in \cite{Rubtsov} ($\varepsilon=0.50$) 
from an analysis of the AGASA data.}
\label{fig:ratio}
\end{figure}

The predicted ratio (\ref{obs}) is shown in Fig.~\ref{fig:ratio}
as function of the observed energy $E$ for $m_X=1\times 10^{13}$~GeV
and for  various values of $\lambda$. This factor depends both on the
local geomagnetic field and the detector type and varies 
therefore from experiment to experiment. 
In Refs.~\cite{AGASA-gamma,Risse}, the effective $\lambda$ was estimated as 
1.2 -- 1.3 for the AGASA site, while Ref.~\cite{So} estimated
$\lambda=2$ for AUGER.

The analysis of AGASA data at $E \geq 1\times 10^{20}$~eV resulted in 
the following upper limits: 
$\varepsilon_{\gamma} < 0.67$ at $E \geq 1.25 \times 10^{20}$~eV 
at 95\% CL in Ref.~\cite{Risse} and $\varepsilon_{\gamma} < 0.5$
at $E \geq 1.0 \times 10^{20}$~eV at 95\% CL in Ref.~\cite{Rubtsov}.

Three remarks are in order.
\begin{itemize}
\item
The energy calibration of different experiments by the position of the dip in
the energy spectrum~\cite{Berezins05} requires a shift of the AGASA
energies by the factor~0.9 and of the HiRes energies by the factor~1.2. 
These shifts lead to a very good agreement of the AGASA and the HiRes
data. Such a shift decreases further the tension between our
prediction of $\varepsilon_\gamma$ and the upper limits from
Ref.~\cite{Risse,Rubtsov}.  
\item
From the Yakutsk and HiRes data we cannot obtain a prediction for the 
$\gamma/p$ ratio within the SHDM model, instead we obtain only an upper
limit. These data agree with the GZK cutoff and can result only in the 
lower limit on $\tau_X$, and thus in the upper limit on the UHE photon
flux from SHDM. On the other hand, $J_{\rm extr}$
is fixed independently, and thus our calculations give only an upper
limit on $\gamma/p$, which is lower than the curves in 
Fig.~\ref{fig:ratio} obtained using the AGASA data. 
\item
Is it possible to use the combined AGASA and Yakutsk data for the
$\gamma/p$ ratio as constraint on the SHDM model for UHECRs? Such a
combination is possible but requires additional assumptions how this
combination is performed. One possibility is to determine the true
UHECR flux averaging  appropriately the AGASA and Yakutsk data. Then 
the SHDM flux would be lower than the one shown in Fig.~\ref{shdm}. 
Consequently, the predicted $\gamma/p$ ratio would be also reduced,
but could now be compared with the limit $\gamma/p \leq 0.36$
from the combined AGASA-Yakutsk data~\cite{Rubtsov}. Another way to
include the Yakutsk data in our analysis is to assume that the lower
flux measured by Yakutsk is caused by a statistical fluctuation.
This assumption raises the question of the compatibility of the
measured AGASA and Yakutsk fluxes. The fluxes of the two experiments
are already  incompatible at lower energies, $10^{19} - 10^{20}$~eV,
where the event numbers are high. Thus their difference cannot be
explained simply by statistical fluctuations. Normalization by the dip
\cite{Berezins05} results in 
an energy shift of the AGASA and Yakutsk energies by a factor 0.9 and 0.75, 
respectively. After this procedure the fluxes coincide perfectly.
This implies that the energies of two Yakutsk events are
considerably below $1\times 10^{20}$, and hence the upper limit 
$\gamma/p$ increases compared to the one given in Ref.~\cite{Rubtsov}.
\end{itemize}

We thus conclude that the obtained upper limits on $\varepsilon_{\gamma}$
(see Fig.~\ref{fig:ratio}) do not exclude, but disfavor the
SHDM model as explanation for the UHECRs.

\section{Anisotropy}
\label{sec:anisotropy}

The anisotropy in the direction to the GC is the most reliable
prediction of the SHDM model. Here we shall present the detailed
calculations of this anisotropy in the form convenient for an analysis
of the Auger data.  

The angular dependence of the UHECR flux from SHDM is given by
\beq
J_{\rm shdm}(\theta)=\frac{1}{4\pi}\int_0^{r_{\rm max}(\theta)}
dr\; \dot{n}_X(R) \,,
\label{flux-gen}
\eeq
where $r$ and $R$ are the distances from the Sun and the GC, respectively, and
$\dot{n}_X$ is the rate of $X$ particle decays given by $n_X(R)/\tau_X$.
As distributions of the DM in the galactic halo we use the 
NFW \cite{NFW} and Moore \etal\ \cite{Moore} profiles, 
\beq
n_X(R)=\frac{n_0}{(R/R_s)^\alpha(1+R/R_s)^{3-\alpha}} ,
\label{n_X}
\eeq
with $\alpha=1$  and  $1.5$ for the NFW and Moore \etal\ profile, 
respectively. We use $R_s=45$~kpc as obtained in Ref.~\cite{BM}.
The distance to the boundary of halo in the direction $\theta$
is given by
\beq
r_{\rm max}(\theta)=r_{\odot}\cos\theta + \sqrt{R_h^2-r_{\odot}^2\sin^2\theta},
\label{r_max}
\eeq
where $R_h=100$~kpc is the size of the DM halo and $r_\odot=8.5$~kpc
the distance of the Sun to the GC.

\begin{widetext}
Changing variable $r \to R$ in the integral of Eq.~(\ref{flux-gen})
we obtain as convenient formula for numerical computations
\beq
J_{\rm shdm}(\theta)=\frac{1}{4\pi \tau_X}\left [
2\int_{r_{\odot}\sin\theta}^{r_{\odot}}dR\; R \frac{n_X(R)}
{\sqrt{R^2-r_{\odot}^2\sin^2\theta}}+
\int_{r_{\odot}}^{R_h}dR\; R \frac{n_X(R)}
{\sqrt{R^2-r_{\odot}^2\sin^2\theta}} \right ].
\label{J_shdm}
\eeq
We define the anisotropy $A$ as the ratio of the flux in the direction of
the GC within the solid angle $\Omega$ to the flux at the same energy
in the perpendicular direction, 
\beq
A(\theta,E)=\frac{J_{\gamma}^{\rm shdm}(\leq\theta,\lambda E)+
J_p^{\rm shdm}(\leq\theta,E)+J_{\rm extr}(E)\Omega(\theta)}
{[J_{\gamma}^{\rm shdm}(90^{\circ},\lambda E)+J_p^{\rm shdm}(90^{\circ},E)+
J_{\rm extr}(E)]\Omega(\theta)} \, .
\label{A}
\eeq
\end{widetext}
Here, $J_{\rm shdm}(\leq\theta)$ is the SHDM flux within the angle $\theta$
relative to the direction to the GC, i.e. within the solid angle 
$\Omega(\theta)=2\pi(1-\cos\theta)$, and $J_{\rm extr}$ is the
extragalactic proton flux taken at the same energy as $J_p^{\rm shdm}$.
Explicitly, the angular dependence of the fluxes of Eq.~(\ref{A}) is
given by
\beq
J_{\rm shdm}(\leq\theta)= \int_0^{\theta}2\pi\sin\theta d\theta 
J_{\rm shdm}(\theta),
\label{J_1}
\eeq
\beq
J_{\rm shdm}(90^{\circ})=\frac{1}{4\pi\tau_X}\int_{r_{\odot}}^{R_h} dR R
\frac{n_X(R)}{\sqrt{R^2-r_{\odot}^2}}.
\label{J_2}
\eeq
The normalized energy-dependent flux $J_{\rm shdm}$ in Eq.~(\ref{J_shdm}) 
is obtained by normalization to AGASA excess according to Eq.~(\ref{AGASA}) 
at $E \geq 1\times 10^{20}$~eV.

\begin{figure}[ht]
\includegraphics[width=0.45\textwidth]{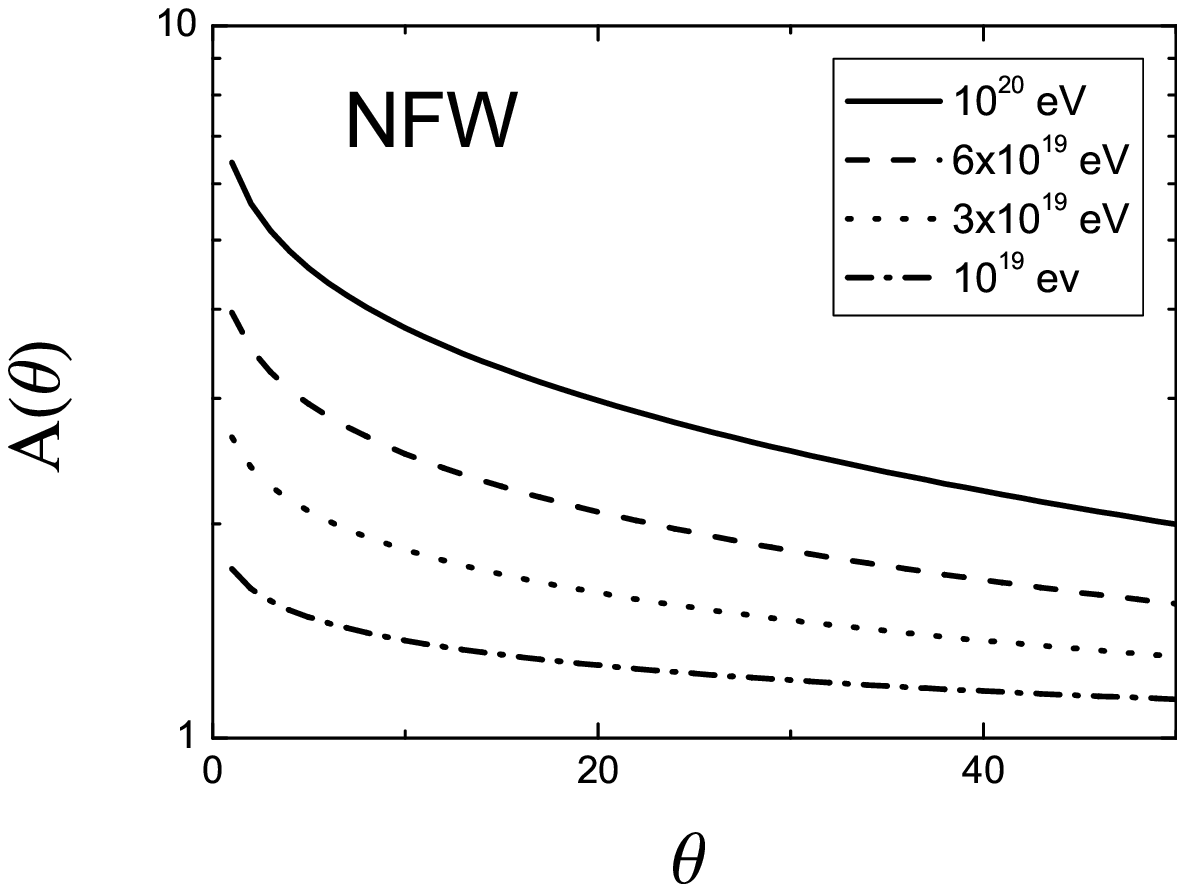}
\includegraphics[width=0.45\textwidth]{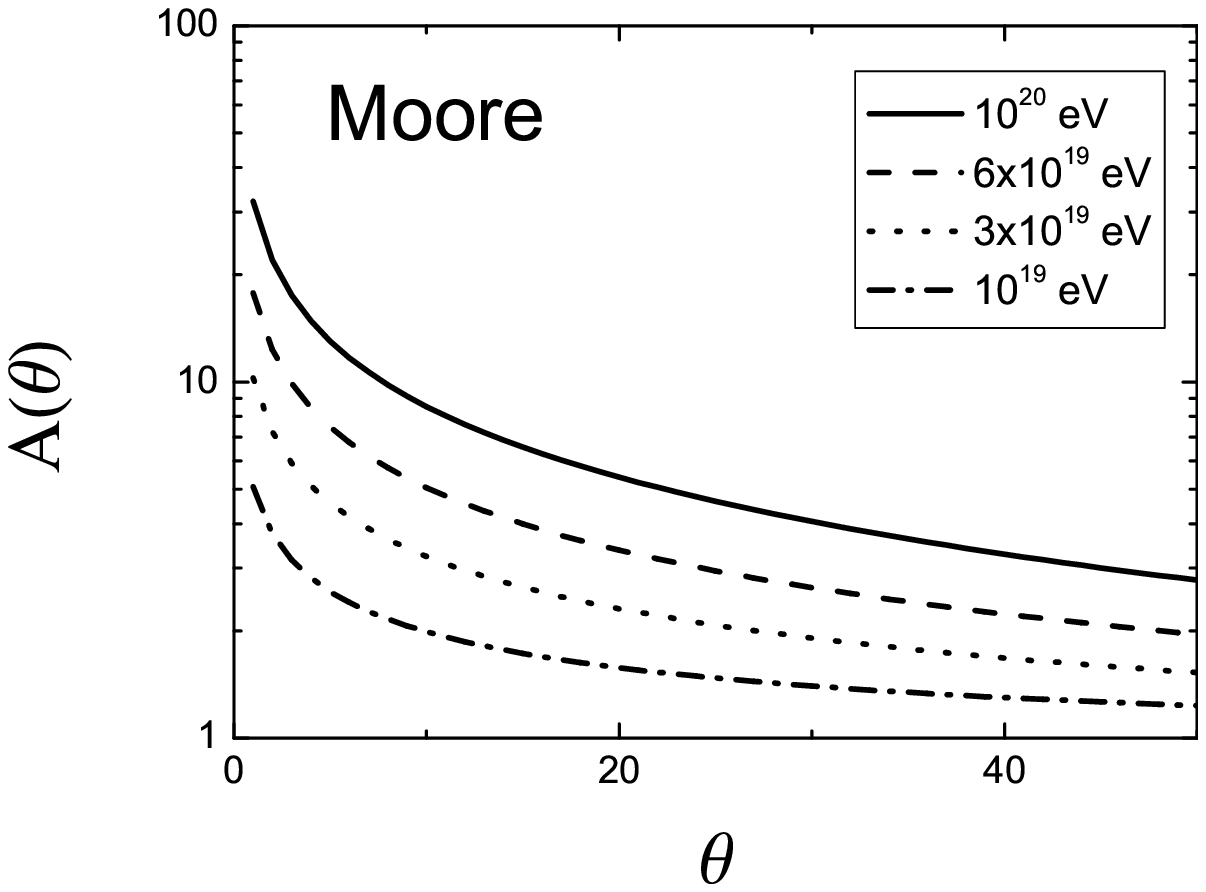}
\caption{Anisotropy $A(\theta,E)$ as defined by Eq.~(\ref{A}) for the NFW 
density profile (top) and the Moore density profile (bottom); 
in both cases $\lambda=2.0$ is used appropriate for the surface
detectors of Auger~\cite{So}.} 
\label{anisotropy}
\end{figure}

Graphical results of our numerical computations
for $m_X=1\times 10^{13}$~GeV are presented in
Fig.~\ref{anisotropy} for the NFW and Moore \etal\ profiles,
while in the Tables \ref{tableNFW} and \ref{tableMoore} numerical
values of the anisotropy are given for small $\theta$ and
different energies. The anisotropy $A(\theta,E)$ weakly 
depends on $\lambda$ and is maximized for 
small $\theta$, but the need for a sufficient number of events will
make the choice of an intermediate value of $\theta$ more suitable. 

\begin{table}
\caption{Numerical values of $A(\theta,E)$ in the case of the NFW DM density profile.}
\label{tableNFW}
\vskip 0.5cm
\begin{tabular}{|c|c|c|c|c|c|c|c|c|c|c|}
\hline
 E | $\theta$ & 1$^\circ$ & 2$^\circ$ & 3$^\circ$ & 4$^\circ$ & 5$^\circ$ & 6$^\circ$ 
& 7$^\circ$ & 8$^\circ$ & 9$^\circ$ & 10$^\circ$ \\
\hline 
1 EeV & 1.03 & 1.03 & 1.02 & 1.02 & 1.02 & 1.02 & 1.02 & 1.02 & 1.02 & 1.01 \\
\hline 
10 EeV & 1.73 & 1.62 & 1.56 & 1.51 & 1.48 & 1.45 & 1.43 & 1.41 & 1.39 & 1.37 \\
\hline 
30 EeV & 2.64 & 2.40 & 2.26 & 2.16 & 2.08 & 2.02 & 1.96 & 1.92 & 1.87 & 1.84 \\
\hline 
60 EeV & 3.96 & 3.52 & 3.26 & 3.08 & 2.94 & 2.83 & 2.73 & 2.65 & 2.57 & 2.51 \\
\hline 
100 EeV &  6.42 & 5.62 & 5.15 & 4.82 & 4.56 & 4.35 & 4.17 & 4.02 & 3.89 & 3.76  \\
\hline
\end{tabular}
\end{table}

\begin{table}
\caption{Numerical values of $A(\theta,E)$ in the case of the Moore DM
 density profile.}
\label{tableMoore}
\vskip 0.5cm
\begin{tabular}{|c|c|c|c|c|c|c|c|c|c|c|}
\hline
 E | $\theta$ & 1$^\circ$ & 2$^\circ$ & 3$^\circ$ & 4$^\circ$ & 5$^\circ$ & 6$^\circ$ 
& 7$^\circ$ & 8$^\circ$ & 9$^\circ$ & 10$^\circ$ \\
\hline 
1 EeV & 1.17 & 1.11 & 1.09 & 1.08 & 1.07 & 1.06 & 1.05 & 1.05 & 1.04 & 1.04 \\
\hline 
10 EeV & 5.08 & 3.75 & 3.16 & 2.81 & 2.57 & 2.40 & 2.26 & 2.15 & 2.06 & 1.99 \\
\hline 
30 EeV & 10.3 & 7.2 & 5.90 & 5.11 & 4.57 & 4.17 & 3.86 & 3.61 & 3.41 & 3.24 \\
\hline 
60 EeV & 17.8 & 12.3 & 9.88 & 8.45 & 7.47 & 6.75 & 6.19 & 5.74 & 5.37 & 5.05 \\
\hline  
100 EeV & 32.1 & 22.0 & 17.5 & 14.8 & 13.0 & 11.7 & 10.6 & 9.80 & 9.11 & 8.53 \\
\hline
\end{tabular} 
\end{table}

\section{conclusions}
\label{sec:conclusions}

Superheavy particles are an interesting candidate for the dark matter in the
universe. They are naturally produced in the expanding universe via
gravitational interactions, when the Hubble parameter $H(t)$ 
exceeds their mass, $H(t) \gsim m_X$. 
The observed density of DM, $\Omega_m=0.27$, determines the mass of
the particle as $m_X \sim (\rm a~few)\times 10^{13}$~GeV. This makes
SHDM a natural candidate for the production of UHECR. 

The SHDM particles ($X$ particles) can be stable (due to, e.g., a discrete 
gauge symmetry) or quasi-stable (due to super-weak discrete 
gauge symmetry breaking). The energy spectrum of produced particles is
approximately a power-law, $\propto E^{-1.9}$. The dominant primary
particles are neutrinos and photons. The only free parameter of the
SHDM model as explanation of the ``AGASA excess'' is the lifetime of
the $X$ particles, which is determined from the UHECR flux as
observed by AGASA as $\tau_X\approx 10^{20}$~years. 

SHDM is accumulated in the halo of our galaxy with an overdensity of 
$2.1\times 10^{5}$ and the produced UHECR flux thus do not has a GZK cutoff. 
The production spectrum $\propto E^{-1.9}$ can explain only the ``AGASA
excess'' at $E \gsim 8\times 10^{19}$~eV. The two other signatures of this 
model are the dominance of photons and the anisotropy towards the
galactic center.  

SHDM as a model for UHECR is at present disfavored by the following points:
\begin{itemize}
\item
The ``AGASA excess'' is not confirmed by the HiRes, Fly's Eye and Yakutsk data,
although the statistics of all these three experiments is too low to
state a serious contradiction. The ``AGASA excess" can be a combined
effect of a small systematic error in the energy determination and of
the low statistics \cite{DBO,Berezins05}.
\item 
Among 17 events with energy $E \gsim 1\times 10^{20}$~eV detected
by all arrays, there is not a single event established as a
gamma-induced air shower. 
\end{itemize}
In this paper we calculated the anisotropy, which can be reliably tested 
by the future data of the Auger experiment. 

The SHDM model has been tuned to explain the ``AGASA excess''. 
If this excess is not be confirmed, or the predicted anisotropy is not
found, it will not exclude SHDM as explanation of the DM in the 
universe. Excluding SHDM as explanation of the observed UHECRs will
put only a lower 
limit on $\tau_X$. The production of UHE photons and neutrinos by DM
in the halo as a subdominant (for the observed UHECR) effect will remain 
a signature of this model. 

\section*{Acknowledgments}
We are grateful to P. Blasi, V. Dokuchaev and Yu. Eroshenko for
valuable discussions and M. Risse for helpful comments. We thank 
ILIAS-TARI for access to the LNGS 
research infrastructure and for financial
support through EU contract RII-CT-2004-506222.


\end{document}